\documentclass[psfig,epsfig,twocolumn,prb,aps,amssymb]{revtex4}
\usepackage{bm,graphicx}
\newcommand{\be}{\begin{equation}}
\newcommand{\ee}{\end{equation}}
\newcommand{\ba}{\begin{eqnarray}}
\newcommand{\ea}{\end{eqnarray}}
\newcommand{\bkt}[1]{\langle #1 \rangle}

\newcommand{\mb}[1]{\ensuremath{\mathbf{#1}}}

\begin{document}

\title{Composite-fermionization of bosons in rapidly rotating atomic traps}
\author{Chia-Chen Chang,$^*$ Nicolas Regnault,$^\dagger$ Thierry Jolicoeur,$^\dagger$  
and Jainendra K. Jain$^*$}
\affiliation{$^*$ Department of Physics, 104 Davey Laboratory,
The Pennsylvania State University, Pennsylvania 16802, U.S.A.}
\affiliation{$^\dagger$ LPA-ENS, D{\'e}partement de Physique, 24, rue 
Lhomond, 75005 Paris, France}
\date{\today}

\begin{abstract}
The non-perturbative effect of interaction can sometimes make interacting
bosons behave as though they were free fermions.
The system of neutral bosons in a rapidly rotating atomic trap is
equivalent to charged bosons coupled to a magnetic field, which
has opened up the possibility of fractional quantum Hall effect 
for bosons interacting with a short range interaction.  Motivated 
by the composite fermion theory of the fractional Hall effect of 
electrons, we test the idea that the interacting bosons
map into non-interacting spinless fermions carrying one
vortex each, by comparing wave functions incorporating this
physics with exact wave functions available for systems containing up to 12 bosons.
We study here the analogy between interacting bosons at filling
factors $\nu=n/(n+1)$ with non-interacting fermions at $\nu^*=n$
for the ground state as well as the low-energy excited states
and find that it provides a good account of the behavior for small $n$,
but interactions between fermions become increasingly important with $n$.
At $\nu=1$, which is obtained in the limit $n\rightarrow \infty$,
the fermionization appears to overcompensate for the repulsive interaction
between bosons, producing an {\em attractive} interactions between
fermions, as evidenced by a pairing of fermions here.
\end{abstract}

\maketitle

\section{Introduction}

The experimental realization of Bose-Einstein
condensation (BEC) of atomic gases\cite{BEC,Ketterle} has generated a rich
variety of phenomena.  In particular, it has allowed the possibility of
testing the remarkable concept of ``statistical transmutation," namely
the idea that interacting bosons may sometimes behave like spinless
fermions.  For contact interactions, it may seem rather sensible for
bosons to emulate fermions, to the extent allowed by symmetry requirements,
because the Pauli principle itself fully takes care of the 
repulsion.  Of course, a conceptual
understanding of how this precisely happens, what it means, and how bosons
can behave like fermions while satisfying the constraints of bosonic exchange symmetry
requires a detailed theory.  The tendency for fermionization
has been appreciated for quite some time for bosons in one 
dimension.\cite{Girardeau60,Lieb63} Girardeau\cite{Girardeau60} showed that for 
an infinitely strong delta function repulsion, the bosonic ground
state wave function $\Psi^B$ is related to the Slater determinant ground
state wave function $\Psi^F$ for spinless fermions in one dimension as:
\be
\Psi^B=|\Psi^F|
\label{zero}
\ee
The problem was solved exactly for an arbitrary strength of the 
interaction by Lieb and Liniger;\cite{Lieb63}
the fermions description is a useful starting point in the strong-coupling limit,
when the interaction strength
is large compared to the Fermi energy.  Recent experiments\cite{Kinoshita04,Parades04}
are in excellent agreement with the Lieb-Liniger theory in the entire range of 
interaction strength, which can be varied in an optically confined one dimensional 
boson system by controlling the density and the confinement strength.

This work is concerned with the possibility of an emergence of 
fermion-like structures in a bosonic system 
in {\em two} dimensions, under conditions 
appropriate for a fractional quantum Hall effect (FQHE) of
bosons.  The familiar FQHE occurs when
charged electrons are confined to two dimensions and exposed to
a strong magnetic field.\cite{Tsui}  There is no realizable system of charged
bosons where FQHE can be studied.  However,
a system of neutral atoms in a rotating trap is mathematically
equivalent to a system of charged bosons in a magnetic field, which,
with confinement to two dimensions, should create,
for sufficiently rapid rotation, a FQHE state of bosons.
BEC systems confined to two dimensions have been created,\cite{Gorlitz}
and their properties have been studied under rotation,
although the FQHE conditions have so far not been achieved.
Rotation of a BEC produces vortices in the condensate.\cite{Matthews,Madison,AboShaeer}
As the rotation frequency is increased, the BEC state is
destroyed and, eventually, the FQHE state may be achieved 
(the latter has no off-diagonal long range order).
These advances have motivated a number of studies of
the FQHE of bosons interacting via a short-range 
interaction.\cite{RotatingBEC,Cooper99,Viefers,Wilkin00,Cooper01,Manninen,Sinova,Schweikhard04,Regnault}

We will assume below that the Landau level (LL) spacing for bosons is 
sufficiently large that it is a good approximation to 
restrict the bosons to the lowest Landau level.  The bosonic 
system is then always in the strong coupling limit, because the 
nature of the state is completely determined by the interaction.
In fact, the solutions are independent of the strength of the interaction, 
which merely sets the energy scale.  It is natural to appeal to the 
fractional quantum Hall effect of electrons for guidance.  
Laughlin's wave function\cite{Laughlin83} can be generalized for  
the ground state at the bosonic filling $\nu=1/2$:
\be
\Psi^B_{\nu=1/2}=\prod_{j<k}(z_j-z_k)^2 \exp\left[-\frac{1}{4}\sum_i|z_i|^2\right] 
\label{Laughlin}
\ee
where $z_j=x_j-iy_j$ denotes the position of the $j$th boson on the two-dimensional
plane, and the magnetic length has been set to unity. 
More generally, the understanding of the electronic FQHE 
is based on the formation of quasiparticles known as composite fermions (CFs);
specifically, the sequence of fractional filling factors $\nu_e=n/(2n+1)$ 
($\nu_e$ refers to the electronic filling factor)
is understood as the integral sequence $\nu^*=n$
of composite fermions.\cite{Jain,Heinonen}  
Application of completely analogous ideas raises 
the possibility that interacting bosons at $\nu=n/(n+1)$ may behave like free
fermions at $\nu^*=n$.  
Jain's wave functions\cite{Jain} can be generalized to  
\be
\Psi^B_{\nu}={\cal P}_{LLL} \prod_{j<k}(z_j-z_k) \Phi^F_{\nu^*}
\label{one}
\ee
where $\Phi^F_{\nu^*}$ is the wave function for non-interacting fermions 
(at the effective filling factor), and ${\cal P}_{LLL}$ projects the 
wave function into the lowest Landau level.  An explicit mapping between 
interacting bosons and noninteracting fermions should be noted.
Eq.~(\ref{one}) produces wave functions for the ground and 
excited states at arbitrary filling in the range $1\geq \nu \geq 1/2$.  
This paper examines their accuracy by comparison with exact wave functions.
If valid, a simplification of the problem is achieved 
through a mapping of a non-trivial interacting boson problem 
into a more amenable non-interacting fermion problem, and many 
essential properties of bosons in rapidly rotating traps 
should find an explanation in terms of almost free particles.

In addition, at filling factor $\nu=1$,
we will consider Moore and Read's Pfaffian wave function\cite{Moore91}, 
given by 
\be
\Psi_{Pf}^B=\text{Pf}\left\{\frac{1}{z_j-z_k}\right\}\prod_{j<k}(z_j-z_k).
  \label{Pfaffian1}
\ee
$\text{Pf}\{M_{jk}\}$ is the Pfaffian of an antisymmetric matrix $M$ with
elements $M_{jk}$, defined as (up to an overall constant)
\be
\text{Pf}\{M_{jk}\}=\sum_\sigma \text{sgn}(\sigma)
M_{\sigma(1)\sigma(2)}M_{\sigma(3)\sigma(4)}\ldots M_{\sigma(N-1)\sigma(N)},
\ee
where the sum is over all permutations $\sigma$, sgn$(\sigma)$ is $+1$ or $-1$ 
depending on whether the permutation is even or odd, and $N$ is an even 
integer.  The Pfaffian has the same form as the projection of the real space 
Bardeen-Cooper-Schrieffer wave function into a fixed number of particles $N$,
and therefore represents a paired state of fermions, as noted by 
Greiter, Wen and Wilczek.\cite{Greiter91} 
The fermion pairing manifests through an incompressible state of bosons.

The mapping into fermions for the bosonic FQHE problem
is conceptually distinct from that applicable 
in one dimension (Eq.~(\ref{zero})).  The modulus of the fermion wave 
function is a manifestly bad approximation for the 
former, because such a wave function has substantial 
mixing with higher Landau levels, and therefore 
a very high kinetic energy.

Much work has already been done toward testing the
composite fermion theory for interacting bosons in a magnetic field.
Many studies take bosons to be in a plane, confined to a disk by
a parabolic confinement; these are analogous to the CF theory of  
electrons confined to a parabolic quantum dot.\cite{Dev92,Jain95}
Viefers, Hansson and Reimann,\cite{Viefers} Cooper and Wilkin\cite{Cooper99},
Wilkin, and Gunn\cite{Wilkin00}, and Manninen {\em et al.}
\cite{Manninen}  have found high overlaps between  
the exact solutions and Jain's wave functions for up to $N=10$ 
particles at the ``magic" angular momenta 
of the {\em yrast} spectrum; further, they also found that the 
state at $\nu=1$ is well described by Moore-Read's wave function.

While a parabolic potential appears naturally for optically 
confined bosonic systems, the strength of confinement can be 
varied, and it may be useful to consider 
the situation without confinement. 
For a large number of bosons, it is natural, in the
simplest approximation, to neglect the
effect of boundaries and concentrate on the bulk properties.
That is most conveniently accomplished in theory by studying bosons in the
spherical geometry,\cite{Haldane83} in which the bosons move on the
two-dimensional surface of the sphere, with a radial magnetic field
produced by a magnetic monopole at the center.  
Exact diagonalization studies have been carried out in the spherical
geometry.  Regnault and Jolicoeur~\cite{Regnault} have shown
that the ground state at $\nu=n/(n+1)$ is
incompressible, consistent with the analogy to filled LL state at 
$\nu^*=n$.  Their results also show evidence of incompressibility at $\nu=1$.
These authors~\cite{Regnault} and 
Nakajima and Ueda\cite{Japan} have studied the excitation spectrum at $\nu=1/2$.
Xie {\em et al.}\cite{Xie91} had earlier studied charged bosons in the
lowest LL; we will briefly consider bosons with long range Coulomb
interaction at the end, but our focus will be on bosons interacting
with a short-range interaction, as appropriate for the atomic
system.  None of these studies, however, has carried out 
a microscopic comparison of the exact eigenstates with 
the wave functions of Eq.~(\ref{one}).
Another possible geometry without boundaries is the 
toroidal geometry, which has been employed by Cooper, Wilkin and 
Gunn~\cite{Cooper01} in the context of bosons in rotating traps.

We will consider below the spherical geometry and
report on detailed and quantitative
tests of the validity of the correspondence between interacting
bosons in the FQHE regime and free fermions in the integral quantum 
Hall regime, which makes definite predictions
for the quantum numbers of the low-energy states of the interacting
boson system, their energies and their eigenfunctions.
Various trial wave functions will be compared with the exact eigenstates and the
predicted energies with the exact eigenenergies.  It is well known that 
Laughlin's wave function,\cite{Laughlin83} which is also a special case of 
Eq.~(\ref{one}), is the exact
solution for the ground state at $\nu=1/2$ for bosons in the lowest
Landau level interacting with a short range interaction.  However,
that by itself does not imply a correspondence between interacting bosons
and free fermions; for that purpose it is necessary to verify 
the correspondence of Eq.~(\ref{one}) for the ground states and 
excitations in a broader range of filling factors.  We will test it  
for the ground state and excitations at $\nu=1/2$, $\nu=2/3$ and $\nu=3/4$.

\section{The Hamiltonian}

We consider a system of $N$ bosons with mass $m$ in a harmonic
trap that is rotating with frequency $\omega$.
In the rotating frame of reference, the system is described by the
Hamiltonian\cite{Dalfovo,Leggett}
\ba
 {\cal H} &=& \sum_{i=1}^N\left\{ \frac{1}{2m} 
(\mb{p}_i-m\omega\hat{\mb{z}}\times\mb{r}_i)^2\right. \nonumber\\
          & &  + \left.\frac{m}{2}\left[(\omega_r^2-\omega^2)(x_i^2+y_i^2)+
\omega_z^2 z_i^2\right] \right\}  \nonumber\\
          & &  + \sum_{i<j}^N V(\mb{r}_i-\mb{r}_j),
\ea
where $\omega_r$ and $\omega_z$ are the radial and axial trap frequencies
respectively. Vectors $\mb{r}_i=(x_i,y_i,z_i)$ represent particle positions.
In an ultra-cold dilute Bose gas, the scattering between particles is
dominated by the $s$-wave scattering process.
It is then an excellent approximation to describe the interaction
by a delta function\cite{Leggett}
\be
V(\mb{r})=\frac{4\pi\hbar^2 a_s}{m}\delta^{(3)}(\mb{r}),
\ee
where $a_s$ is the $s$-wave scattering length, assumed to be positive
in this work.
When $\omega$ and $\omega_r$ are identical, the Hamiltonian resembles that of
particles with charge $e$ in a magnetic field $\mb{B}=(2m\omega/e)\hat{\mb{z}}$.
An effective magnetic length is defined as $\ell=\sqrt{\hbar/(2m\omega)}$.
The effective cyclotron frequency is defined as $\omega_c=eB/m=2\omega$.
If the axial trap is strong enough such that the wave function along the
$z$ direction is the ground state of
the harmonic potential in the $z$ axis,
the system enters a two dimensional (2D) regime where the potential
felt by particles is written as\cite{Regnault}
\be
 V(\mb{r})=g\,\delta^{(2)}(\mb{r}),
 \label{interaction}
\ee
with $g=\hbar^2 a_s \sqrt{8\pi}/(m\ell_z)$, where $\ell_z=\sqrt{\hbar/(m\omega_z)}$.
The energy scale in the 2D regime
is set by the effective coupling constant $g$.
We will assume below that the interaction strength is sufficiently 
small compared to the Landau level spacing that LL mixing is 
negligible.  From our experience with electronic FQHE, we know 
that a modest amount of LL mixing does not significantly alter the results.

\section{Composite fermion theory}

For bosons in the lowest Landau level, there are three situations.   
(i) For $\nu<1/2$, there are many linearly independent 
wave functions that vanish upon coincidence of bosons, producing  
an enormous ground state degeneracy. 
(ii) For $\nu=1/2$ there is a single wave function that 
has zero energy for the delta function interaction, giving a 
non-degenerate ground state here.  It remains the ground state 
for arbitrarily high coupling $g$, and may be considered to be 
the analog of the Girardeau wave function of the one dimensional problem.
(iii) For the excitations at $\nu=1/2$, or for any eigenfunctions 
at $\nu>1/2$, there are no wave functions in the 
lowest Landau level that vanish when two particles coincide. 
While no exact results are available here, analogy to fermions 
gives plausible wave functions that we now describe.

\subsection{$\nu=1/2$}

For $\nu=1/2$, Laughlin's wave function for the ground state is
given by (in the spherical geometry)
\be
\Psi^B_{1/2}=\Phi_1^{2},
\ee
where
\be
\Phi_1=\prod_{j<k}(u_jv_k-v_ju_k),
\ee
is the wave function of the lowest filled Landau level.
\be
(u_j,v_j)=(\cos(\theta_j/2) e^{i\phi_j/2},\sin(\theta_j/2) e^{-i\phi_j/2})
\ee
are the spinor coordinates describing the position of a particle on
the surface of a sphere.
It is the exact ground state for bosons at $\nu=1/2$
interacting with a delta function interaction, which can be
seen straightforwardly by noting that $\Phi_1^2$ is the only wave function
at $\nu=1/2$ that is confined to the lowest LL and has zero interaction
energy for the delta function interaction.
The wave functions $\Phi_1^{2p}$ with $p\geq 2$ are not
relevant for the short range interaction, as these are degenerate with a
large number of other states.

\subsection{$\nu\geq 1/2$}

For {\em electrons} in the lowest LL, the CF theory\cite{Jain} hypothesizes that
strongly interacting electrons map into weakly interacting fermions of
a new kind, called composite fermions.  The composite fermions 
experience an effective magnetic field given by
$B^*=B-m\rho\phi_0$, where $B$ is the external magnetic field,
$\phi_0= hc/e$, and $m$ is an even integer.
Equivalently, the filling factor of composite fermions, $\nu^*$,
is related to the electron filling factor, $\nu$, by $\nu=\nu^*/(m\nu^*+1)$.
This is interpreted in terms of electrons having captured an even
number ($m$) of flux quanta of the external magnetic field
to become composite fermions, which no longer see the magnetic
flux that they have assimilated into themselves.  This physics suggests 
the wave functions 
$\Psi^F_{\nu}={\cal P}_{LLL}\Phi_1^{m}\Phi_{\nu^*}$, where
$\Phi_{\nu^*}$ is the Slater determinant wave function for
non-interacting electrons at $\nu^*$, $\Phi_1$ is the wave function of one filled
Landau level, and ${\cal P}_{LLL}$ projects the wave function into the
lowest Landau level, as appropriate for very large magnetic fields.
These wave function explicitly relate the eigenfunctions of
interacting electrons at $\nu$ to those of non-interacting fermions at $\nu^*$,
and have been tested both in the
spherical geometry\cite{Jain97} and the disk geometry.\cite{Dev92,Jain95,Jeon04}

The considerations in the preceding paragraph are 
readily generalized to bosons by taking $m$ to be an odd
integer.  We specialize to $m=1$ (other odd integer values not
being relevant to the problem of our interest) and filling factors
\be
\nu=\frac{n}{n+1}.
\ee
which correspond to $\nu^*=n$ of fermions.
The wave function at $\nu$ is now given by  
\be
\Psi^B_{\nu}={\cal P}_{LLL}\Phi_1\Phi_{n}.
\label{psib}
\ee
which is the spherical analog of Eq.~(\ref{one}).
The Jastrow factor $\Phi_1$ now attaches a single
vortex to each fermion in $\Phi_{n}$.

These two equations define the mapping between interacting
bosons and non-interacting fermions in microscopic detail.
The first equation has implications about the structure of the
low-energy eigenstates of the interacting boson system, whereas the
last gives trial wave functions for the eigenstates, and also the
eigenenergies.  There are two ways to physically think about the above
equations.  (i) Bosons have captured an odd
number of vortices each to convert into a composite fermion.\cite{Leinaas}  
(ii) The bosons are represented as bound states of
fermions and an odd number of vortices.  The tests
below, of course, are independent of the interpretation.

We note that the ground state and excitations of interacting
bosons at $\nu=n/(n+1)$ are images of the ground state and
excitations of fermions at $\nu^*=n$ according to Eq.~(\ref{psib}).
The wave function for the ground 
state at $\nu=n/(n+1)$ is
given by Eq.~(\ref{NumericalWf}) with $\Phi_n$ taken as the wave
function of the ground state at $\nu^*=n$, i.e., the $n$ filled
Landau level state.
The wave function for the excited state is similarly related to
the lowest energy particle-hole excitation, i.e. an exciton
at $\nu^*=n$.  The eigenstates of the spherical geometry 
are labeled by the total orbital angular momentum, $L$.  
The ground state has $L=0$, which implies uniform density. 
It has no adjustable parameters, given that the wave function 
of $n$ filled Landau levels is unique.
The wave function for the exciton for any given $L$ 
is also determined completely by group theory, and therefore 
is free of any adjustable parameters.

A subtle feature of the composite fermion theory ought to be noted.
The ground state wave function at $\nu=1/2$ ($\nu^*=1$) is given by
$\Psi^B_{1/2}=\Phi_1^2$
(no lowest-Landau level projection is required here, because
the wave function is already in the lowest Landau level).
It manifestly eliminates spatial coincidence of particles, and thus
has zero interaction energy for the contact interaction potential.
As mentioned earlier, no such wave functions can be written, even 
in principle, for the excited states at $\nu=1/2$ or for any 
states at $\nu>1/2$.  The CF theory circumvents this problem by
first neglecting the lowest LL constraint to write wave functions ($\Phi_1\Phi_n$)
in which bosons do not occupy the same spatial position, and then projecting
them into the lowest Landau level, hoping that this
would capture the actual correlations within the lowest LL.
The wave functions $\Psi^B$ are
in general much more complicated than Laughlin's wave
function at $\nu=1/2$.  Their validity is far from obvious, and their 
confirmation would provide a non-trivial
evidence for the composite-fermionization of the bosonic system.

\subsection{$\nu=1$}

We will also be interested in the nature of the state
in the limit of $n\rightarrow \infty$, i.e. at $\nu=1$.  Let
us recall what happens for electrons in this limit, which
corresponds to $\nu_e=1/2$ for electrons.  If the residual interactions
between composite fermions are negligible, a Fermi sea of
composite fermions is obtained here (the state with an infinite
number of filled Landau levels is another representation of a
Fermi sea), as proposed by Halperin, Lee, and Read.\cite{Halperin93}
That provides a good description of the
compressible state at $\nu_e=1/2$, also explaining why there
is no FQHE here.\cite{FermiSea}   However, in the second
Landau level, electrons form an {\em incompressible} state when
the Landau level is half full (which corresponds to a total filling
of $\nu_e=5/2$), which appears to be best described by 
Moore-Read's wave function.
This implies that the mapping into {\em non}interacting
composite fermions is no longer valid, and one must consider the residual
interactions between them, which presumably cause a pairing instability
of the CF Fermi sea.\cite{Scarola}

If bosons behaved like {\em non-}interacting fermions in the limit
of $n\rightarrow \infty$, the system at $\nu=1$ would be
analogous to the Halperin-Lee-Read Fermi sea.  On the other hand, 
if the bosons map into interacting fermions,
Moore-Read's wave function becomes a plausible candidate.
In the spherical geometry, it is given by
\be
  \Psi^B_{Pf}=\text{Pf}\left\{\frac{1}{u_jv_k-u_jv_k}\right\}\prod_{j<k}^N(u_jv_k-u_jv_k).
  \label{Pfaffian}
\ee

\section{Calculation}

We will study the wave function in Eq.~(\ref{psib}) for the ground 
states and excitations at $\nu=1/2$, $\nu=2/3$  and $\nu=3/4$.  For 
technical convenience, we will {\em define} the lowest LL projection as follows:
\be
\Psi^B_{n/(n+1)}=\Phi_1^{-1}{\cal P}_{LLL}\Phi_1^2\Phi_{n}
\label{NumericalWf}
\ee
The philosophy of ``lowest LL projection"
is to obtain from the unprojected wave function $\Phi_1\Phi_n$,
which has some amplitude in higher Landau levels,
a wave function that resides strictly within the lowest LL.
For electrons, at least two ways of accomplishing this have been
useful;\cite{Dev92,Jain97} they both give lowest LL wave functions
that are very
close to the exact eigenstates.  The more convenient of these two
methods\cite{Jain97} relies on having at least two
factors of $\Phi_1$; it does
not work for $\Phi_1\Phi_n$ but requires $\Phi_1^2\Phi_n$.
That is the reason for defining the projection as in Eq.~(\ref{NumericalWf}).
We refer the reader to the literature~\cite{Jain97}
for the explicit
construction of the lowest Landau level projected wave functions
${\cal P}_{LLL}\Phi_1^2\Phi_{n}$, which can be used here without
change.  The presence of $\Phi_1$ in the denominator is not a cause
for concern, because ${\cal P}_{LLL}\Phi_1^2\Phi_{n}$, 
being antisymmetric, also contains
the factor $\Phi_1$ in it.  We have not tested the relative merits
of this method of projection as opposed to a direct projection, but,
based on our experience with fermions, we expect them to produce
more or less the same lowest LL wave function.

To compare $\Psi^B$ with the exact wave functions
$\Psi^{ex}$, we will calculate their overlap:
\be
{\cal O}^2\equiv \frac{|\langle\Psi^{ex}_{n/(n+1)}|\Psi^B_{n/(n+1)}\rangle|^2}
{|\langle\Psi^{ex}_{n/(n+1)}|\Psi^{ex}_{n/(n+1)}
\rangle||\langle\Psi^B_{n/(n+1)}|\Psi^B_{n/(n+1)}\rangle|},
\label{ov}
\ee
For the Metropolis Monte Carlo evaluation, it is convenient to  
rewrite it as 
\be
{\cal O}^2\equiv 
\frac{\left[|\bkt{\Psi^{ex}_{n/(n+1)}|\Psi^B_{n/(n+1}}|/|\bkt{\Psi^B_{n/(n+1)}
|\Psi^B_{n/(n+1)}}|\right]^2}
     {|\bkt{\Psi^{ex}_{n/(n+1)}|\Psi^{ex}_{n/(n+1)}}|/
|\bkt{\Psi^B_{n/(n+1)}|\Psi^B_{n/(n+1)}}|}
\ee
Then, using the wave function $\Psi^B_{n/(n+1)}$ as the sampling weight,
both the numerator and the denominator can be calculated simultaneously.
$\Psi^B_{n/(n+1)}$ represents either the ground state wave function
or the CF exciton wave function at $\nu=n/(n+1)$.
The corresponding exact wave functions are obtained 
by the L\'anczos algorithm.

Another measure of the quantitative accuracy of the CF
description is the comparison between the predicted energy with
the exact one.  The CF prediction for the
energy of the ground or excited states is given by
\be E =
\frac{\langle\Psi^B_{n/(n+1)}|V|\Psi^B_{n/(n+1)}\rangle}
{\langle\Psi^B_{n/(n+1)}|\Psi^B_{n/(n+1)}\rangle}.
\label{montecarlo}
\ee
Even though the wave functions are rather complicated, the
integral can be evaluated by the Metropolis Monte Carlo method.
We find it convenient to write the numerator as
\ba
& & \sum_{i<j}\langle\Psi^B_{n/(n+1)}|\delta^{(2)}
(\bm{\Omega}_i - \bm{\Omega}_j)|\Psi^B_{n/(n+1)}\rangle \nonumber \\
&=& \frac{N(N-1)}{2}\int \prod_{i=1}^N d\bm{\Omega}_i
    \,\delta^{(2)}(\bm{\Omega}_1-\bm{\Omega}_2)|\Psi^B_{n/(n+1)}
(\bm{\Omega}_1,\bm{\Omega}_2,\ldots)|^2 \nonumber \\
&=& \frac 1 2 \frac{N(N-1)}{4\pi R^2}\int \prod_{i=1}^N\,d\bm{\Omega}_i
    |\Psi^B_{n/(n+1)}(\bm{\Omega}_2,\bm{\Omega}_2,\ldots)|^2.
    \label{energy}
\ea
where, in the spherical geometry, the unit vector
$\bm{\Omega}_i=(\sin\theta_i\cos\varphi_i,\sin\theta_i\sin\varphi_i,\cos\theta_i)$
describes the position of particles on the surface of the sphere,
and we have used $\int d\bm{\Omega}_1/4\pi R^2=1$ in the last step
($R=$ the radius of the sphere), which expresses the integral in a form where
$|\Psi^B_{n/(n+1)}(\{\bm{\Omega}_i\})|^2$ can be used as the sampling
function.

For the Monte Carlo evaluation of the overlap,
occupation number basis states are transformed
into real space basis wave functions, which are permanents.
The permanent is an analog of a determinant of a square matrix $M$
with elements $M_{jk}$ in which all signs are taken as positive
in the expansion of minors. In general, it can be written as
$\text{per}(M_{jk})=\sum_\sigma\prod_{j=1}^N M_{j,\sigma(j)}$,
where $\sigma$ are permutations of $N$ indices.
We evaluate the permanents using the Ryser algorithm.\cite{Knuth}
Typically we perform $10^3\sim 10^5$ iterations in one
Monte Carlo run. For larger systems, the majority
of the computational time is spent on evaluating
permanents. For example, we need to evaluate
61108 permanents at each Monte Carlo step for
a system of $N=12$ bosons at $\nu=3/4$ which takes
approximately 480 CPU hours for $10^3$ iterations on a single node of a 
PC cluster, consisting of dual 2.4GHz PentiumIV processors,
to accumulate the desired accuracy. We use as many as 10 nodes
to increase the efficiency.

In the energy calculation, the wave function
in Eq.~(\ref{NumericalWf}) consists of a linear
combination of several determinants for an excited state
at a given angular momentum $L$.  (For the ground state, 
only one determinant needs to be evaluated.)
The calculation for energy is far less time consuming than
that for the overlap.  We perform about $1.2\times 10^7$
iterations in a single Monte Carlo run. The quoted statistical 
uncertainty in the calculation reflects one standard deviation
from 10 independent runs.  To give an idea of the computation time,
approximately 40 CPU hours are needed for the
ground state energy of $\nu=3/4$ at $N=12$.  In Eq.~(\ref{energy}), 
the positions of the first two particles are identical.
To avoid numerical division by zero, 
we set $\mb{\Omega}_1=\mb{\Omega}_2+\delta\mb{\Omega}$.
The results are independent of $|\delta\mb{\Omega}|$ 
provided it is sufficiently small; 
we typically use $|\delta\mb{\Omega}|=10^{-6}$.

The Moore-Read wave function is known to be the exact ground state for a 
three body interaction \cite{Greiter91,Read99}
\ba
\cal{H}_{{\rm pfaff}}&=&\sum_{i<j<k} 
\delta^{(2)}\left(\mb{r}_i-\mb{r}_j\right)\delta^{(2)}\left(\mb{r}_i-\mb{r}_k\right).
\ea
It can therefore be obtained by exact diagonalization using L\'anczos 
algorithm in the spherical geometry.  That provides a direct method to 
evaluate the scalar product involved in the overlap calculation.

\section{Quantitative comparisons}

\begin{figure}
\includegraphics[angle=-90,scale=0.32]{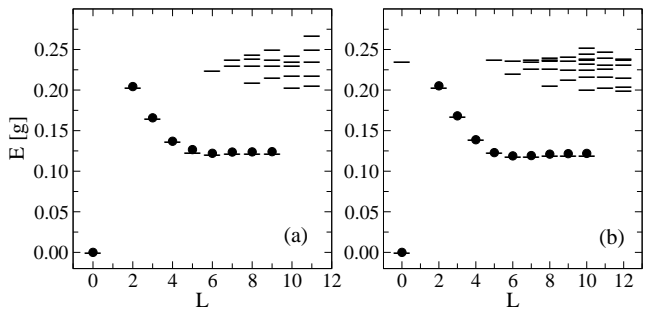}
\caption{The low-energy spectrum of (a) $N=9$ and (b) $N=10$ interacting 
bosons at $\nu=1/2$, interacting with a delta function interaction with 
strength $g$.  Dashes represent exact results, while the 
dots show predictions of the composite fermion theory.
Spherical geometry is used in the calculation, and $L$ is the total 
orbital angular momentum.  In this and the subsequent two figures, the 
statistical uncertainty from Monte Carlo sampling (not 
shown) is smaller than the symbol size.}
\end{figure}

\vspace{1cm}

\begin{figure}
\includegraphics[angle=-90,scale=0.32]{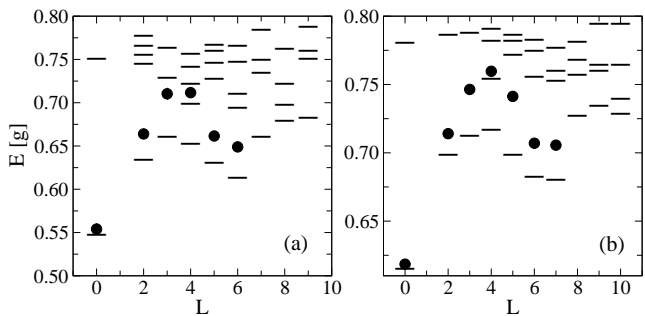}
\caption{The low-energy spectrum of 
(a) $N=10$ and (b) $N=12$ interacting bosons at $\nu=2/3$.
Various symbols have the same meanings as in Fig.~1.
}
\end{figure}

\begin{figure}
\includegraphics[angle=-90,scale=0.31]{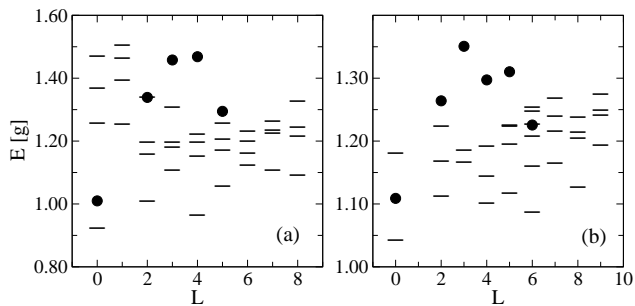}
\caption{The low-energy spectrum 
of (a) $N=9$ and (b) $N=12$ interacting bosons
at $\nu=3/4$.  See the caption of Fig.~1 for the definition of 
various symbols.  
}
\end{figure}

The results of our study are summarized in Table I and Figs.~1-3.
The Table I gives the overlaps of exact eigenstates $\Psi^{ex}$ with the
wave functions $\Psi^B$ for the ground state and first excited state.
We make the following remarks.

(i) For $\nu=1/2$, $2/3$, and $3/4$, the structure of the low-energy 
states is in clear correspondence with that of fermions at $\nu^*=1$, 2, 
and 3.  In all these cases, the ground state is a uniform state 
($L=0$), well separated from the other states by a gap, as 
shown earlier.\cite{Regnault} The quantum numbers of the low-energy 
excitations at $\nu=1/2$ and $2/3$ can also be understood by analogy 
to $\nu^*=1$ and 2, although the correspondence for excitations 
is poor for $\nu=3/4$.    
Thus, an inspection of the low-energy spectrum of interacting bosons 
at $\nu$ clearly shows similarity with fermions at $\nu^*$.

(ii) At $\nu=1/2$, the wave function $\Phi_1^2$ is known to be exact.
Our calculations of the overlap and energy for this filling
constitute a non-trivial test of the correctness of our computer codes.

\begin{table*}
\begin{ruledtabular}
\begin{tabular}{cccccccccccccccccc}
$\nu $ & $N$ & ${\cal O}_{\text{gr}}^2$ & $L$ & ${\cal O}_{\text{ex}}^2$ & $\nu$ & $N$ & ${\cal O}_{\text{gr}}^2$ & $L$ & ${\cal O}_{\text{ex}}^2$ & $\nu$ & $N$ & ${\cal O}_{\text{gr}}^2$ & $L$ & ${\cal O}_{\text{ex}}^2$ & $\nu$ & $N$ & ${\cal O}_{\text{gr}}^2$ \\ \hline
  1/2 & 4 & 1.0000 & 4 & 0.9972 & 2/3 &  4 & 1.0000     & 2 & 1.0000    & 3/4 &  9 & 0.8084(73) & 4 & 0.5613(48)& 1  &  4 & 1.00000 \\
      & 5 & 1.0000 & 4 & 0.9965 &     &  6 & 0.9850     & 4 & 0.7544(05)&     & 12 & 0.735(84)  & 6 & 0.480(62) &    &  6 & 0.97279 \\
      & 6 & 1.0000 & 5 & 0.9959 &     &  8 & 0.9820(10) & 5 & 0.8701(14)&     &    &            &   &           &    &  8 & 0.96687 \\
      & 7 & 1.0000 & 5 & 0.9954 &     & 10 & 0.9724(89) & 6 & 0.855(12) &     &    &            &   &           &    & 10 & 0.95922 \\
      & 8 & 1.0000 & 6 & 0.9945 &     &    &            & &             &     &    &            &   &           &    & 12 & 0.88435 \\
      & 9 & 1.000  & 6 & 0.9954 (2) &     &    &            & &             &     &    &            &   &           &    & 14 & 0.88580 \\
\end{tabular}
\end{ruledtabular}
\caption{The overlap of the exact wave 
functions for the ground state and the first excited state at $\nu=1/2$, $2/3$, $3/4$, and $1$ 
with the trial wave functions of Laughlin ($\nu=1/2$ ground state),
Moore and Read ($\nu=1$ ground state), and Jain (other states), 
for several particle numbers $N$.  ${\cal O}_{\text{gr}}$ is the overlap for 
the ground state, and ${\cal O}_{\text{ex}}$ for the first excited 
state, which occurs at the orbital angular momentum $L$.  
The definition of the overlap is given in Eq.~(\ref{ov}). 
The statistical uncertainty in the last two digits is shown in parentheses 
when it is larger than $10^{-5}$. 
For $\nu=1$, only the ground state overlap is shown, which has been evaluated exactly. 
}
\end{table*}

\begin{table*}
\begin{ruledtabular}
\begin{tabular}{cccccccccccc}
 $\nu $ & $N$ & ${\cal O}_{\text{gr}}^2$ & $\nu$ &  $N$ & ${\cal O}_{\text{gr}}^2$ & $\nu$ & $N$ & ${\cal O}_{\text{gr}}^2$ & $\nu$ & $N$ & ${\cal O}_{\text{gr}}^2$ \\ \hline
  1/2   &  4  &  0.9999 &  2/3  &  4   &  1.0000     &  3/4  &  9  &   0.8163(76) &  1  &  4  & 1.00000 \\
        &  5  &  0.9998 &       &  6   &  0.9901     &       & 12  &   0.820(41)  &     &  6  & 0.97279 \\
        &  6  &  0.9997 &       &  8   &  0.9898(02) &       &     &              &     &  8  & 0.97710 \\
        &  7  &  0.9997 &       & 10   &  0.9870(11) &       &     &              &     & 10  & 0.96589 \\
        &  8  &  0.9994 &       &      &             &       &     &              &     & 12  & 0.91645 \\
        &  9  &  0.9994 &       &      &             &       &     &              &     & 14  & 0.92133 \\
\end{tabular}
\end{ruledtabular}
\caption{The overlap of the {\em Coulomb} ground state wave function 
at $\nu=1/2$, $2/3$, $3/4$, and $1$
with the trial wave functions of Laughlin ($\nu=1/2$ ground state),
Moore and Read ($\nu=1$ ground state), and Jain ($\nu=2/3$ and $3/4$ ground states), 
for several particle numbers $N$.  (Table I dealt with the exact wave functions for  
a short range interaction.) The statistical uncertainty in the last two digits is shown when 
it is larger than $10^{-5}$.  The overlaps for $\nu=1$ (last 
column) are exact.
}
\end{table*}

(iii) The low-energy excited states at $\nu=1/2$ are extremely well
described, quantitatively, as excitons of composite 
fermions.\cite{Dev92b,Scarola00}  The composite-fermion
theory predicts that there is a single state at orbital angular
momenta from $L=2$ to $L=N$, which is clearly the case in exact
diagonalization studies.  (At $\nu^*=1$ there is also an exciton
at $L=1$, but its wave function is annihilated upon projection into
the lowest Landau level.\cite{Dev92b})  Further, the energy of the CF exciton
is in excellent agreement with the exact energy.
Previous studies~\cite{Regnault,Japan} studied the excited states at 
$\nu=1/2$ by exact diagonalization, but did not provide a microscopic 
understanding.

(iv) For $\nu=2/3$, the CF theory  
provides an excellent approximation for the ground state,
with very high overlaps and very good energies for 10 and 12
particles.  The CF theory again correctly predicts the quantum numbers of
the low energy excitations, and also the qualitative shape of the
exciton dispersion curve, but the energies are
now off by up to $\sim$ 50\%.  At $\nu=3/4$, the situation becomes worse.
In accordance with the prediction of the CF theory, the
ground state has $L=0$, but no well defined branch of
excitations may be identified with the CF exciton
branch; furthermore, the energies predicted by the CF theory 
are quite inaccurate, for both the ground and excited
states.  These studies show that the CF description worsens
with increasing $n$ along the sequence $\nu=n/(n+1)$.

(v) At $\nu=1$, a good account of the ground state is obtained 
through analogy to a paired fermion state, as can be seen from
the overlaps given in the last column of the Table I.
This result is consistent with the earlier studies in the 
toroidal and disk geometries\cite{Cooper99,Wilkin00,Cooper01}.

One may ask to what extent the difference between bosons at $\nu=n/(n+1)$ and
electrons at $\nu_e=n/(2n+1)$ has to do with the fact that the bosons
are interacting via a short-range, contact interaction, as opposed to
the long-range Coulomb interaction for the electrons.
To investigate this issue, we obtain exact wave functions for a system
of charged bosons interacting via the Coulomb interaction.
Table II presents their overlaps with various wave functions.  The
CF theory is in better agreement with the Coulomb states at $\nu=n/(n+1)$,
but the overall behavior is qualitatively similar.
For example, the overlaps for $\nu=3/4$ are not high, and much 
smaller than those at $\nu_e=3/7$ for the electron FQHE.
The paired wave function is also a better approximation for 
the Coulomb ground state at $\nu=1$ than it is for the hard-core interaction; 
in contrast, it is not valid for  
the electronic state at the corresponding filling $\nu_e=1/2$.
These observations indicate that both the particle statistics and the
nature of the interaction are responsible for the differences
in the behaviors of fermions and bosons in a magnetic field.

\section{Conclusion}

The above results allow us to make the following conclusions:
(i) The mapping into free fermions is qualitatively valid for a range 
of parameters.  It correctly captures the incompressibility of the ground 
state at $\nu=1/2$, 2/3 and 3/4. 
(ii) The mapping is also quantitatively very accurate for the ground state and 
excitations at $\nu=1/2$ and for the ground state at $\nu=2/3$,  
but becomes progressively worse with increasing $n$.  This implies that
the residual interactions
between fermions become increasingly more important with $n$,
and must be considered for a more complete understanding.
(iii) A qualitative indication of the breakdown  
of the {\em free}-fermion model is the appearance of
a paired state at $\nu=1/2$; {\em non}interacting fermions would have
produced a Fermi sea here.  The residual interactions 
between fermions thus cause a qualitative change in the nature of the 
state beyond certain $n$, presumably through a pairing instability;
we cannot, however, ascertain from our study for what $n$ a phase 
transition occurs.

Overall, these results establish that the mapping into 
strictly free fermions is valid only for a limited range of parameters, but the 
mapping into weakly interacting fermions provides a useful starting point 
for a wider range of parameters.  One might ask why the interacting 
fermion language is to be preferred over the original interacting boson model.
The reason is that the interaction between the fermions is much weaker, 
with a large part of the repulsive interaction taken care of by the 
Pauli avoidance.

The bosonic FQHE should be contrasted with the FQHE of electrons 
along the sequence $\nu_e=n/(2n+1)$, for which the 
mapping into free composite fermions
remains valid for the entire parameter range.
The wave functions of Eq.~(\ref{one}) are 
known to provides an excellent description of the state for 1/3, 2/5,
and 3/7, where exact results are available,\cite{Dev92b,Jain97}
and presumably also for higher $n$, as evidenced by the
experimental observation of many fractions along the sequence
as well as of the Fermi sea of
composite fermions at $\nu=1/2$.\cite{Halperin93,FermiSea}

It is noteworthy that pairing occurs in a model of bosons 
with purely repulsive interactions.  As stressed in Ref.\onlinecite{Scarola},
even when the interaction between the original particles is repulsive, the 
effective interaction between the emergent particles may be attractive.
That appears to be the case at $\nu=1$.  Here, 
bosons dress themselves with vortices to turn into 
fermions, but that presumably overcompensates for the repulsive 
interaction, thereby producing an attractive interaction between the fermions.

\section{Acknowledgment}

JKJ thanks N. Cooper, K. O'Hara, E.H. Rezayi, S. Viefers, and D.S. Weiss for 
useful discussions.  Partial support of this research by the National 
Science Foundation under grant no. DMR-0240458 is acknowledged. 
We are grateful for a computer time allocation of IDRIS-CNRS, and to the
High Performance Computing (HPC) Group led by V. Agarwala, J. Holmes,
and J. Nucciarone, at the Penn State University ASET (Academic Services
and Emerging Technologies) for assistance and computing time with the
LION-XL cluster.

\end{document}